\documentclass[12pt,fleqn]{article}

\def\keywords{\vspace{.3em}
    \if@twocolumn
      \small\it Keywords\/\bf---$\!$%
    \else
      \begin{center}\small\bf Keywords\end{center}\quotation\small
    \fi}
\def\endkeywords{\vspace{0.6em}\par\if@twocolumn\else\endquotation\fi
    \normalsize\rm}

\usepackage{a4wide}
\usepackage{amsmath}
\usepackage{amssymb}
\usepackage{epsfig}
\usepackage{float}
\usepackage{times}
\usepackage{graphicx}
\usepackage{subfigure}
\usepackage{citesort}

 \restylefloat{figure}

\newcommand{\tran}[1]{#1^{\scriptscriptstyle\text{T}}}

\newcommand{\vect}[1]{{\mathbf{#1}}}
\newcommand{\mat}[1]{{\mathbf{#1}}}
\newcommand{\beqna}{\begin{eqnarray}}
\newcommand{\eeqna}{\end{eqnarray}}
\newcommand{\bmat}{\begin{bmatrix}}
\newcommand{\bemat}{\end{bmatrix}}
\newcommand{\nn}{\nonumber}
\newcommand{\eref}[1]{(\ref{#1})}

\renewcommand{\a}{\vect{a}}
\renewcommand{\b}{\vect{b}}

\renewcommand{\d}{\vect{d}}

\newcommand{\m}{\vect{m}}
\newcommand{\n}{\vect{n}}

\renewcommand{\r}{\vect{r}}
\newcommand{\s}{\vect{s}}

\renewcommand{\u}{\vect{u}}

\newcommand{\y}{\vect{y}}
\newcommand{\z}{\vect{z}}

\newcommand{\0}{\vect{0}}
\newcommand{\1}{\vect{1}}

\newcommand{\C}{\mat{C}}

\newcommand{\I}{\mat{I}}

\newcommand{\R}{\mat{R}}
\renewcommand{\S}{\mat{S}}

\begin{document}

\title{Nonlinear MMSE Multiuser Detection Based on \\ Multivariate Gaussian Approximation}
\author{\small Peng Hui Tan, {\em Student Member, IEEE},
Lars K. Rasmussen, {\em Senior Member, IEEE}
\date{February 15, 2005}
\thanks{P. H. Tan and L. K. Rasmussen are with the Department of Computer
Engineering, Chalmers University of Technology, G\"oteborg,
Sweden. L. K. Rasmussen is also with the Institute for
Telecommunications Research, University of South Australia. P. H.
Tan and L. K. Rasmussen are supported in parts by the Swedish
Research Council for Engineering Sciences under grants no.
271-1999-390 and 217-1997-538. P. H. Tan is also supported by the
Personal Computing and Communication (PCC++) Program under Grant
PCC-0201-09, and L. K. Rasmussen is supported by the Australian
Research Council under ARC Grant DP0344856 and by the Australian
Academy of Science International Scientific Collaboration
Program.}}

\maketitle

\vspace{-1.2cm}

\begin{abstract}
In this paper, a class of nonlinear MMSE multiuser detectors are
derived based on a multivariate Gaussian approximation of the
multiple access interference. This approach leads to expressions
identical to those describing the probabilistic data association
(PDA) detector, thus providing an alternative analytical
justification for this structure. A simplification to the PDA
detector based on approximating the covariance matrix of the
multivariate Gaussian distribution is suggested, resulting in a
soft interference cancellation scheme. Corresponding multiuser
soft-input, soft-output detectors delivering extrinsic
log-likelihood ratios are derived for application in iterative
multiuser decoders. Finally, a large system performance analysis
is conducted for the simplified PDA, showing that the bit error
rate performance of this detector can be accurately predicted and
related to the replica method analysis for the optimal detector.
Methods from statistical neuro-dynamics are shown to provide a
closely related alternative large system prediction. Numerical
results demonstrate that for large systems, the bit error rate is
accurately predicted by the analysis and found to be close to
optimal performance.
\end{abstract}


\section{Introduction}
It is well-known that the computational complexity of individually
optimal detection for direct-sequence code-division
multiple-access (CDMA) grows exponentially with the number of
users \cite{Ver98bk}, as the computation of the marginal
posterior-mode (MPM) distribution is required. Maximum {\em a
posteriori} probability (MAP) detection for each user is therefore
far too complex for practical CDMA systems with even a moderate
number of users. The exponentially growing complexity has inspired
a considerable effort in finding low complexity suboptimal
alternatives capable of resolving the detrimental effects of
multiple-access interference (MAI).

Interference cancellation (IC) strategies have been subject to
particular attention due to low complexity, a simple modular
structure and competitive performance \cite{Ras03wiley}. Early
work was focused on linear cancellation and hard decision
cancellation \cite{PatHol94JSAC,Var90TC}. More recently, soft
decision cancellation have been shown to provide performance
improvements. In \cite{Tan01TC} it was shown that soft decision
cancellation based on convex projections provides an iterative
solution to the convex-constrained multiuser maximum-likelihood
problem. The well-known result that the optimal nonlinear minimum
mean squared error (MMSE) estimate is the conditional
posterior-mode mean was used in \cite{Tar95ISIT} for a
decision-feedback receiver. Similar arguments were used in
\cite{Gol99ICASSP} to arrive at a soft decision IC structure, and
the same structure was derived in \cite{Mul98bk} based on neural
networks arguments. Even though this cancellation structure has a
low complexity of order $\mathcal{O}(K^2)$, numerical examples
show that near single-user performance can be achieved for large
systems \cite{Mul98bk}.

In \cite{Luo01CL}, the probabilistic data association (PDA) method
was introduced for multiuser detection as a low complexity
nonlinear alternative. The decision statistics of the users are
modelled as binary random variables where the MAI is approximated
as multivariate Gaussian noise. The {\em a posteriori} probability
(APP) for the data symbols of each user is updated sequentially
given the associated APPs of all other users. Although this scheme
has a low computational complexity of order $\mathcal{O}(K^3)$, it
can achieve near single-user performance for systems with a
moderate number of users \cite{Luo01CL}.

The most celebrated multiuser detectors applied for iterative
multiuser decoding of coded CDMA are based on linear filtering,
e.g.,
\cite{AleGra98ETT,Sti03VT,ShiSch01JSAC,Wan99TC,ElG00JSAC,MarVuc01ISIT,LinRas04AUS,RasGra04IT}.
Parallel IC (PIC) and linear MMSE filtered PIC were investigated
in \cite{AleGra98ETT,Sti03VT,ShiSch01JSAC} and
\cite{Wan99TC,ElG00JSAC}, respectively. In \cite{MarVuc01ISIT}, it
was observed that for low-complexity detectors, information
combining over iterations can be rewarding, providing performance
and system load gains. The partial cancellation structure in
\cite{MarVuc01ISIT} was justified in \cite{LinRas04AUS} as
recursive maximal ratio combining over all previous iterations,
while a more complicated vector Kalman filter applied across
iterations was presented in \cite{RasGra04IT}. Nonlinear multiuser
detectors based on list detection have been developed for
iterative multiuser decoding and shown to provide equally
impressive performance gains at low complexity
\cite{ReiGra02Glob}. As the PDA detector generates APPs directly,
it has been applied for iterative multiuser decoding with only
minor modifications, also demonstrating competitive gains
\cite{Tan03ISIT}.

Large system performance analysis techniques from statistical
mechanics and statistical neuro-dynamics have been applied
successfully for performance analysis of some multiuser detectors.
In \cite{Tan02IT}, the performance of the optimal multiuser
detector was analyzed based on the replica method. This approach
has further been developed in \cite{GuoVer02bk}, and in
\cite{Cai03IT} for coded CDMA. A different approach inspired by
statistical neuro-dynamics was used in \cite{Kab03JPA} to arrive
at a large system analysis for a belief propagation (BP) multiuser
detector. Methods from statistical neuro-dynamics
\cite{Oka95NW,Ama88NN} have also been applied in \cite{Tan03IT}
for large system analysis of PIC.

In this paper, a class of nonlinear MMSE (NMMSE) multiuser
detectors are derived based on a multivariate Gaussian
approximation of the MAI. The computation of the NMMSE estimate
requires a sum of terms, which grows exponentially in numbers with
the number of users. Using the multivariate Gaussian
approximation, this summation is replaced by integration, reducing
the complexity significantly. The expressions describing this
approach is shown to be identical to the description of the PDA
detector in \cite{Luo01CL}, thus providing an alternative
analytical justification.

A simplification to the NMMSE/PDA detector\footnote{In the
remaining of the paper, this detector is referred to as the
simplified PDA detector.}, based on approximating the covariance
matrix of the multivariate Gaussian distribution with a diagonal,
is suggested. The corresponding soft interference cancellation
scheme is similar to the IC structure of the detectors in
\cite{Gol99ICASSP,Mul98bk}and can be implemented in parallel or
serially. The corresponding complexity is of the order of IC,
namely $\mathcal{O}(K^2)$ as compared to the PDA with an order of
complexity of $\mathcal{O}(K^3)$.

Multiuser soft-input, soft-output (SISO) detectors delivering
extrinsic log-likelihood ratios (LLRs) at the output are derived
from the class of NMMSE-based detectors. The multiuser SISO
detectors are applied for iterative multiuser decoding of coded
CDMA and found to converge to single-user performance at loads
larger than linear multiuser SISO alternatives.

Finally, a large system performance analysis is conducted for the
simplified PDA. In the large system limit, the bit error rate
performance of this detector can be accurately predicted and
related to the replica method analysis for the optimal detector
\cite{Tan02IT}. Methods from statistical neuro-dynamics can also
be used for a closely related alternative large system prediction
\cite{Kab03JPA,Oka95NW}. It follows that the simplified PDA has
the same predicted large system performance as the optimal
detector. Numerical results show that for large systems, the bit
error rate (BER) is accurately predicted by the analysis and found
to be close to optimal performance.

The paper is organized as follows. In Section \ref{sec:model}, the
uncoded and coded CDMA discrete-time models are presented together
with the standard iterative multiuser decoding structure. In
Section \ref{sec:NMMSE} nonlinear minimum mean squared error
estimation, leading to the marginal posterior-mode (MPM) decision,
is briefly reviewed providing the setting for the multivariate
Gaussian approximation considered in Section \ref{sec:multivar}.
The simplified PDA is derived in Section \ref{sec:spda}, while the
corresponding NMMSE-based multiuser SISO detectors are detailed in
Section \ref{sec:siso}. The large system analysis of the
simplified PDA is derived in Section \ref{sec:LSP}, numerical
results are presented in Section \ref{sec:NR} and concluding
remarks are summarized in Section \ref{sec:Con}.

\section{System Model} \label{sec:model}
\begin{figure}[htbp]
  \begin{center}
  \includegraphics*[width=140mm]{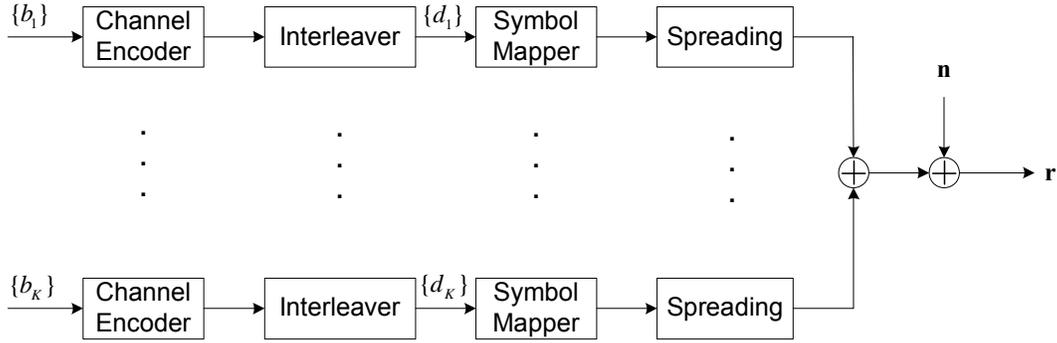}
    \caption{Discrete-time model for coded CDMA.}
    \label{fig:transmit}
  \end{center}
\end{figure}

An elaborate discrete-time system model for CDMA is developed from
first principles in \cite{Ras98bk}. The discrete-time model
described below is a simplified, special case of this general
model. For simplicity, assume a symbol-synchronous CDMA system
with $K$ users, binary data symbols and binary spreading with
processing gain $N$. Random spreading is assumed where each binary
chip is modulated onto a common chip waveform for transmission.
The output of a bank of $K$ chip-matched filters is given by
\beqna\label{eqn:chip_model}
    \r = [\s_1,...,\s_k,...,\s_K]\; \d + \n = \S\d +\n,
\eeqna where $\S\in\{\pm 1/\sqrt{N}\}^{N \times K}$ is the
spreading matrix, $\d\in\{\pm 1\}^K$ is the data symbol vector,
$\n$ is a zero-mean additive white Gaussian noise (AWGN) vector
with covariance matrix $\sigma^2\I$, and $N_0 = 2 \sigma^2$ is the
one-sided spectral density of the white Gaussian noise. The model
is illustrated in Figure \ref{fig:transmit} within the error
control coded model.

Some notation that will prove useful later on. At chip interval
$\mu$, the received signal is described by $r_\mu = \sum_{k=1}^N
s_{\mu k}d_k + n_\mu$, where $r_\mu$, $s_{\mu k}$ and $n_\mu$ are
corresponding elements of the vectors $\r$, $\s_k$ and $\n$,
respectively. In addition, let $\S_k =
[\s_1,...,\s_{k-1},\s_{k+1},...,\s_K]$ be the spreading matrix
with column $k$ removed. The model in \eref{eqn:chip_model} can be
further developed to include bit-level matched filtering as $\y =
\tran{\S} \r = \R \d + \z$, where $\mathsf{E}\{\z \tran{\z} \} =
\sigma^2\R$. It follows that $y_k = \sum_{j=1}^K R_{kj}d_j + z_k$,
where $y_k$ and $z_k$ are respective elements of vectors $\y$ and
$\z$, while $R_{kj}$ is the corresponding element of the matrix
$\R$.

\begin{figure}[htbp]
  \begin{center}
  \includegraphics*[width=150mm]{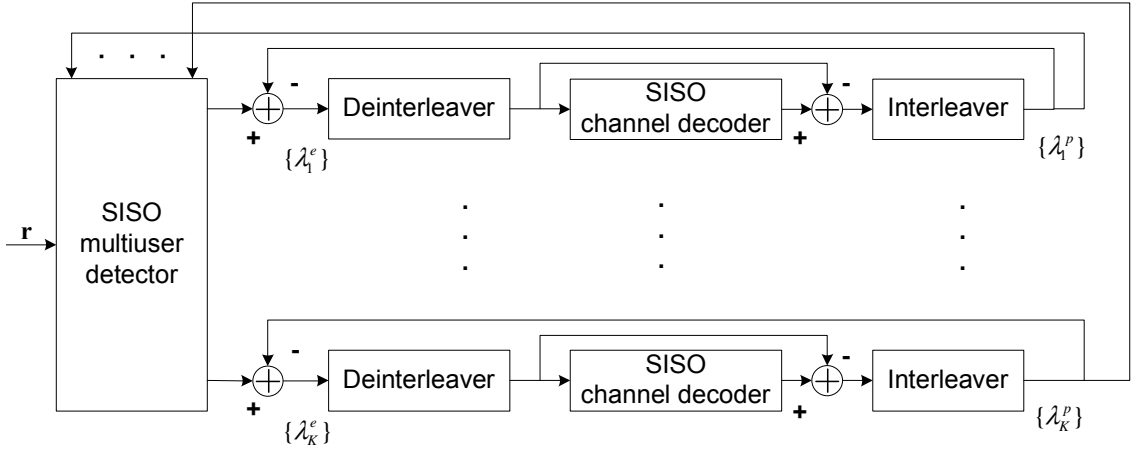}
    \caption{General structure for iterative multiuser decoding.}
    \label{fig:it_mud}
  \end{center}
\end{figure}

When error control coding is introduced, the model is extended as
shown in Figure \ref{fig:transmit}. Now the binary data symbols
are encoded, interleaved and mapped onto a binary phase-shift
keying constellation in order to arrive at the code symbol vector
$\d$, which corresponds to the data symbol vector in the model for
the uncoded case. In this paper, we consider iterative multiuser
decoding for the coded case with the corresponding decoding
structure shown in Figure \ref{fig:it_mud}.
A multiuser SISO detector computes extrinsic LLRs of the code bits
for all the users based on the received signal and \emph{a priori}
LLRs of the code bits. The extrinsic LLRs of user $k$ are
deinterleaved and input to an APP decoder for the error control code
applied by user $k$. This single-user decoder outputs extrinsic
LLRs, which are interleaved and, together with extrinsic LLRs of all
the other users, forwarded to the multiuser SISO as \emph{a priori}
LLRs for the next iteration. This type of iterative multiuser
decoder is a direct application of the turbo decoding principle and
commonly used for iterative multiuser decoding
\cite{Wan99TC,Tan03ISIT,Cai03IT,RasGra04IT}.

\section{Nonlinear MMSE Estimation} \label{sec:NMMSE}
Let the nonlinear MMSE data estimate for user $k$ be denoted as
$m_k = \mathsf{g}^*(d_k,\r)$, where $ \mathsf{g}^*(d_k,\r)$ is the
nonlinear function that minimizes the mean squared error
$\mathsf{E}\{(d_k-\mathsf{g}(d_k,\r))^2\}$. In order to find the
optimal nonlinear function, the mean squared error is expressed as
an expectation of a conditional expected value
$\mathsf{E}\{\mathsf{E}\{(d_k-\mathsf{g}(d_k,\r))^2|\r\}\}$
\cite{Pap84}. Since the inner expectation is always positive, the
minimum is achieved by: \beqna\label{eqn:nmmse}
    \min_{\mathsf{g}(d_k,\r) \in \mathbb{G}}
    \mathsf{E}\{[d_k-\mathsf{g}(d_k,\r)]^2|\r\} =
    \min_{\mathsf{g}(d_k,\r) \in \mathbb{G}} \sum_{d_k=\pm 1}
[d_k-\mathsf{g}(d_k,\r)]^2 \mathsf{Pr}(d_k|\r), \eeqna where
$\mathbb{G}$ is the relevant set of nonlinear functions. The
solution is the conditional mean
$\mathsf{E}\{\mathsf{Pr}(d_k|\r)\}$ \cite{Pap84}, leading to
\beqna\label{eqn:mapp} m_k = \mathsf{g}^*(d_k,\r) = \sum_{d_k=\pm
1} d_k \mathsf{Pr}(d_k|\r) = \sum_{\d \in \{-1,+1\}^K}d_k
\mathsf{Pr}(\d|\r). \eeqna Note that the polarity of $m_k$ in eqn.
\eref{eqn:mapp} is in fact the marginal posterior-mode decision,
i.e., \beqna
    d_k^* = \arg\,\,\max_{d_k=\pm 1} \mathsf{Pr}(d_k|\r)
= \mathsf{sign}\left\{\sum_{\d \in \{-1,+1\}^K}d_k
\mathsf{Pr}(\d|\r)\right\}.\nn \eeqna

Based on eqns. \eref{eqn:nmmse} and \eref{eqn:mapp}, the NMMSE
data estimates for all the users can be described by a set of $K$
optimization problems: \beqna m_k = \arg \min_{\tilde{m}_k \in
\mathbb{R}}\sum_{d_k}(d_k-\tilde{m}_k)^2\mathsf{Pr}(d_k|\r),
\hspace{0.5cm} \text{for } k=1,2,...,K, \nn \eeqna where $m_k$ is
the NMMSE data estimate for user $k$. The $K$ problems can be
solved independently since $\mathsf{Pr}(d_k|\r)$ can be computed
independently for each user.

Following Bayes' rule, the marginal posterior-mode distribution
can be found as \beqna\label{eqn:mmpd} \mathsf{Pr}(d_k|\r)
=\frac{\mathsf{Pr}(d_k)\mathsf{p}(\r|d_k)}{\sum_{d_k}\mathsf{Pr}(d_k)
\mathsf{p}(\r|d_k)}. \eeqna Here, the probability density function
(pdf) $\mathsf{p}(\r|d_k)$ is found as a sum over $2^{K-1}$ terms
as follows: \beqna\label{eqn:prdk} \mathsf{p}(\r|d_k) =
\sum_{\d\backslash d_k \in \{-1, +1 \}^{K-1}}
\mathsf{p}(\r|\d)\mathsf{Pr}(\d \backslash d_k), \eeqna where
$\d\backslash d_k$ denotes a vector containing all the elements in
$\d$ except $d_k$. This approach is however impractical for large
system loads, as the computational complexity grows exponentially
with the number of users. As an alternative, a multivariate
Gaussian approximation is introduced below.

\section{Multivariate Gaussian Approximation}\label{sec:multivar}
Consider the received signal at chip level. The conditional pdf at
chip interval $\mu$ is \beqna
\mathsf{p}(r_\mu|\d)=\frac{\mathsf{exp}\left[-\frac{1}{2\sigma^2}
\left(r_\mu-s_{\mu k}d_k - \Delta_{\mu
k}\right)^2\right]}{\sqrt{2\pi\sigma^2}}, \nn \eeqna where
$\Delta_{\mu k}= \sum_{l\neq k}s_{\mu l}d_l$ is the corresponding
MAI. The conditional symbol-level pdf in \eref{eqn:prdk} can then
be expressed as \beqna \mathsf{p}(\r|d_k) &=& \sum_{\d\backslash
d_k \in \{-1, +1 \}^{K-1}} \prod_{\mu=1}^N
\mathsf{p}(r_{\mu}|\d)\mathsf{Pr}(\d \backslash
d_k) \nn \\
&=& \sum_{\d\backslash d_k \in \{-1, +1 \}^{K-1}}\mathsf{Pr}(\d
\backslash d_k) \frac{\mathsf{exp}\left[-\frac{1}{2\sigma^2}
\left\|\r-\s_{k}d_k -
\Delta_{k}\right\|^2\right]}{(2\pi\sigma^2)^{N/2}},\label{eqn:prdk_chip}
\eeqna where $\mat{\boldsymbol{\Delta}}_{k} =
\tran{[\Delta_{1k},...,\Delta_{Nk}]}$ is a vector for user $k$,
containing the MAI contributions for each chip interval.

To reduce complexity, the probability distribution function of the
random variable vector $\mat{\boldsymbol{\Delta}}_{k}$ is
approximated by a multivariate Gaussian pdf. The summation in
\eref{eqn:prdk_chip} can thus be replaced by an $N$-fold
integration over the support of $\mat{\boldsymbol{\Delta}}_{k}$
\beqna\label{eqn:aprdk} \mathsf{p}(\r|d_k) \approx
\int_{-\infty}^{\infty} \ldots \int_{-\infty}^{\infty}
\mathsf{p}(\r,\mat{\boldsymbol{\Delta}}_{k}
|d_k)\,d\mat{\boldsymbol{\Delta}}_{k} = \int_{-\infty}^{\infty}
\ldots \int_{-\infty}^{\infty} \prod_{\mu=1}^N
\mathsf{p}(r_\mu|\Delta_{\mu
k},d_k)\mathsf{p}(\mat{\boldsymbol{\Delta}}_{k})
\,d\mat{\boldsymbol{\Delta}}_{k}, \eeqna where
$d\mat{\boldsymbol{\Delta}}_{k} = \prod_{\mu =1}^N d\Delta_{\mu
k}$ denotes differentials for integration. The multivariate
Gaussian pdf is described as follows. Since $\Delta_{\mu k} =
\sum_{l\neq k} s_{\mu l}d_l$, it is reasonable to assume that the
corresponding mean and covariance are \beqna u_{\mu k} =
\mathsf{E}\left\{\Delta_{\mu k}\right\}= \sum_{l\neq k} s_{\mu
l}m_l \nn \eeqna and
\begin{eqnarray}\label{eqn:cov1}
    \mathsf{Cov}\left\{\Delta_{\mu k}\Delta_{\nu k}\right\}&=&
\mathsf{E}\left\{\Delta_{\mu k}\Delta_{\nu k}\right\} -
\mathsf{E}\left\{\Delta_{\mu k}\right\}
\mathsf{E}\left\{\Delta_{\nu k}\right\} \nn\\
&=& \sum_{j\neq k}s_{\mu j}s_{\nu j}(1-m_j^2) + \sum_{j\neq k}
\,\sum_{l \neq j,k}s_{\mu j}s_{\nu l}\left(\mathsf{E}
\left\{d_jd_l\right\}-m_jm_l\right).
\end{eqnarray}
In the second term in \eref{eqn:cov1}, the expectation $\mathsf{E}
\left\{d_jd_l\right\}$ must be computed. This computation has a
complexity of the order of $\mathcal{O}(K^2)$. To reduce
complexity, the second term is omitted in the following. As $K$
grows large, it is expected that $\mathsf{E} \left\{d_jd_l\right\}
\rightarrow m_jm_l$ and thus, the second term becomes negligible.
The effect of removing this term is considered in Section
\ref{sec:NR} using numerical examples. With this simplification,
the covariance matrix of $\mat{\boldsymbol{\Delta}}_{k}$ is
reduced to \beqna \mat{\boldsymbol{\Omega}}_{k} =
\mathsf{Cov}\{\mat{\boldsymbol{\Delta}}_{k}
\tran{\mat{\boldsymbol{\Delta}}_{k}}\} =  \sum_{l\neq k} (1-m_l^2)
\s_l\tran{\s_l} =
\S_k\mathsf{Diag}[\1-\m_k\circ\m_k]\tran{\S_k},\nn \eeqna where
$\mathsf{Cov}\{\Delta_{\mu k}\Delta_{\nu k}\} = \sum_{l\neq k}
s_{\mu l}s_{\nu l}(1-m_l^2)$, $\m_k =
\tran{[m_{1},m_{2},...,m_{k-1},m_{k+1},...,m_{N}]}$ and $\a \circ
\b$ denotes the Hadamard-product \cite{Hor85bk} of vectors $\a$
and $\b$, respectively. The multivariate Gaussian pdf of
$\mat{\boldsymbol{\Delta}}_{k}$ is then \beqna
\mathsf{p}(\mat{\boldsymbol{\Delta}}_{k})=\frac{\mathsf{exp}
\left[-\frac{1}{2}\tran{(\mat{\boldsymbol{\Delta}}_{k} -
\u_{k})}\mat{\boldsymbol{\Omega}}_{k}^{-1}
(\mat{\boldsymbol{\Delta}}_{k} - \u_{k})\right]}{(2\pi)^{N/2}
\sqrt{\mathsf{det}[\mat{\boldsymbol{\Omega}}_{k}]}}, \nn\eeqna
where $\u_k =\tran{[u_{1k},u_{2k},...,u_{Nk}]} = \S_k \m_k$.

Substituting this into \eref{eqn:aprdk} and performing the
$N$-fold integration yields \beqna
\mathsf{p}(\r|d_k)&\varpropto&\exp\left\{-\frac{1}{2}
\tran{\left(\r-\s_kd_k-\u_k\right)}
(\mat{\boldsymbol{\Omega}}_{k}+\sigma^2\I)^{-1}
\left(\r-\s_kd_k-\u_k\right)\right\}\nn\\
&\varpropto&\exp\left\{d_k\tran{\left(\r-\u_k\right)}
\C_k^{-1}\s_k\right\} = \mathsf{exp}\left\{d_k \tran{\s}_k
\C_k^{-1} \left(\r-\S_k \m_k \right) \right\},\label{eqn:cpdf1}
\eeqna where $\C_k=\mat{\boldsymbol{\Omega}}_{k}+\sigma^2\I$.
It follows that the NMMSE estimate is given by \beqna m_k &=&
\sum_{d_k = \pm 1} d_k \mathsf{Pr}(d_k|\r) = \sum_{d_k = \pm 1}
d_k \frac{\mathsf{Pr}(d_k)
\mathsf{p}(\r|d_k)}{\sum_{d_k}\mathsf{Pr}(d_k)
\mathsf{p}(\r|d_k)} \nn \\
&=& \mathsf{tanh}\left[\lambda_k^p/2+\tran{\s}_k \C_k^{-1}
\left(\r-\S_k \m_k \right)\right],\label{eqn:pda1} \eeqna where
$\lambda_k^p =
\mathsf{log}[\mathsf{Pr}(d_k=1)/\mathsf{Pr}(d_k=-1)]$ is the
\emph{a priori} log-likelihood ratio (LLR).

The above detector is know as the PDA detector, first suggested in
\cite{Luo01CL}. Our contribution is to relate the PDA detector to
the NMMSE estimation problem, which shows that the corresponding
output is an approximation to the conditional \emph{a posteriori}
mean. Also, it is clear from \eref{eqn:pda1} that the PDA detector
corresponds to a nonlinear, filtered IC structure.

Solving the nonlinear system of equations in \eref{eqn:pda1}
requires a computational complexity of the order of
${\mathcal{O}}(K^3)$ \cite{Luo01CL}, where the complexity is
dominated by the inversion of $\C_k$. A simplified approach is
suggested below, approximating $\C_k$ with a diagonal matrix.

\section{Simplified Probabilistic Data Association
Detection}\label{sec:spda} For large systems,  the diagonal
elements of $\C_k$ are dominant, encouraging the following
approximation $\C_k = (\mat{\boldsymbol{\Omega}}_{k}+\sigma^2\I)
\approx (\sigma_k^2+\sigma^2)\I$, where $\sigma_k^2=\alpha(1-Q)$
with $\alpha=K/N$ being the system load and $Q = (1/K)\sum_k
m_k^2$. The conditional pdf \eref{eqn:cpdf1} is then simplified to
\beqna\label{eqn:spdf} \mathsf{p}(\r|d_k) &=& \prod_{\mu=1}^N
\frac{\mathsf{exp}\left[-\frac{(r_\mu-s_{\mu k}d_k - u_{\mu
k})^2}{2(\sigma_{k}^2+\sigma^2)} \right]
}{\sqrt{2\pi(\sigma_{k}^2+\sigma^2)}} =
\frac{\mathsf{exp}\left[-\frac{\left\|\r-\s_{k}d_k - \S_k
\m_k\right\|^2}{2(\sigma_{k}^2+\sigma^2)} \right]
}{\left[(2\pi(\sigma_{k}^2+\sigma^2)\right]^{N/2}} \nn \\
&\varpropto& \mathsf{exp}\left[\frac{d_k \tran{\s}_k \left(\r-\S_k
\m_k\right)}{\sigma_{k}^2+\sigma^2}  \right], \eeqna which leads
to \beqna\label{eqn:mkspda} m_k =
\mathsf{tanh}\left[\frac{\lambda_k^p}{2} + \frac{\tran{\s}_k
\left(\r- \S_k \m_k\right)}{\sigma_{k}^2+\sigma^2}\right]=
\mathsf{tanh}\left[\frac{\lambda_k^p}{2} + \frac{y_k- \sum_{j\neq
k} R_{k j} m_j}{\sigma_{k}^2+\sigma^2}\right]. \eeqna Note that
\eref{eqn:mkspda} is similar to the iterative soft-decision
multi-stage interference cancellation (MIC) scheme suggested
independently in \cite{Tar95ISIT,Mul98bk,Gol99ICASSP}. The MIC is
described by \beqna\label{eqn:mkic} m_k
=\mathsf{tanh}\left[\frac{\lambda_k^p}{2} + \frac{y_k -
\sum_{j\neq k} R_{kj}m_j}{\sigma^2+\sum_{j\neq k}
R_{kj}^2(1-m_j^2)} \right]. \eeqna For large $K$ and $N$, the term
$\sum_{j\neq k} R_{kj}^2(1-m_j^2)$ is well approximated by
$\alpha(1-Q)$, using the fact that
$\mathsf{E}\left\{R_{kj}^2\right\}= 1/N$.

A simple way to solve \eref{eqn:mkspda} is by iteration over all
users from an initial solution $\m^0$. This can be done in
parallel as \beqna\label{eqn:pspda} m_k^{t+1} = \omega m_k^{t} +
(1-\omega)\mathsf{tanh}\left[ \frac{\lambda_k^p}{2} + \frac{y_k -
\sum_{j\neq k} R_{kj}m_j^t}{\sigma^2+\alpha(1-Q^t)} \right],
\eeqna where superscript $t$ denotes the corresponding variable at
iteration $t$. Also, $0\le\omega<1$ is a weighting factor which
improves the convergence properties of the parallel iteration in
\eref{eqn:pspda}. Similar weighting factor approaches have been
applied to linear cancellation and convex-constrained cancellation
in \cite{GraSch01TC,Tan01TC}.

The fixed-point problem in \eref{eqn:mkspda} can also be solved
with a serial iteration as \beqna\label{eqn:sspda} m_k^{t+1} =
\omega m_k^{t} +
(1-\omega)\mathsf{tanh}\left[\frac{\lambda_k^p}{2} + \frac{y_k -
\sum_{j=1}^{k-1} R_{kj}m_j^{t+1}-\sum_{j=k+1}^{K}
R_{kj}m_j^{t}}{\sigma^2+\alpha\left\{1-\frac{1}{K}
\left[\sum_{j=1}^{k-1}(m_j^{t+1})^2 +
\sum_{j=k}^{K}(m_j^{t})^2\right]\right\}} \right]. \eeqna It
should be noted that convergence is not assured in general.
However, for a series of numerical experiments, it has been
observed that the serial implementation with $\omega = 0$ always
converged while a nonzero weighting factor is required for the
parallel case to ensure convergence.

In the following, the parallel implementation in \eref{eqn:pspda}
is denoted as the parallel simplified PDA (PSPDA) and the serial
implementation in \eref{eqn:sspda} is denoted as the serial
simplified PDA (SSPDA).

\section{Multiuser Decoding}\label{sec:siso}
The multiuser detectors considered in this paper are based
directly on estimating the marginal-mode probability distribution
function. This feature makes these detectors well suited for
low-complexity iterative multiuser decoding, requiring only minor
modifications. Based on the general iterative multiuser decoding
approach in \cite{Wan99TC,Tan03ISIT,Cai03IT}, the extrinsic LLRs
of the detectors developed above are derived.

>From \eref{eqn:mmpd}, the LLR for user $k$ based on the marginal
mode probability distribution is \beqna \Lambda^{\text{APP}}_k =
\mathsf{log}\frac{\mathsf{Pr}(d_k=1|\r)}{\mathsf{Pr}(d_k=-1|\r)} =
\mathsf{log}
\frac{\mathsf{Pr}(d_k=1)\mathsf{p}(\r|d_k=1)}{\mathsf{Pr}(d_k=-1)
\mathsf{p}(\r|d_k=-1)}= \lambda^p_k + \lambda^e_k, \nn\eeqna where
$\lambda^p_k$ is the \emph{a priori} LLR and $\lambda^{e}_k =
\mathsf{log} \frac{\mathsf{p}(\r|d_k=1)}{\mathsf{p}(\r|d_k=-1)}$
is the extrinsic LLR for user $k$. A multiuser SISO based on the
PDA detector is determined from \eref{eqn:cpdf1}. The
corresponding LLR is $\lambda^{e}_k = 2 \tran{\s}_k \C^{-1}_k
\left(\r - \S_k \m_k \right)$, following a sufficient number of
iterations of the PDA detector, according to \eref{eqn:pda1}
either in parallel or serially. This is to arrive at as good an
approximation as possible to the conditional \emph{a posteriori}
mean. Considering the approximate conditional pdf in
\eref{eqn:spdf}, the corresponding LLR for a multiuser SISO based
on the simplified PDA detector is $\lambda^{e}_k = \frac{2
\tran{\s}_k \left(\r - \S_k \m_k \right)}{\sigma_k^2+\sigma^2}$,
again assuming sufficient iterations of \eref{eqn:pspda} or
\eref{eqn:sspda} to get a good approximation to $\m_k$ for all
$k$.

Note that we now have two separate iterations, namely the overall
multiuser decoding iteration, exchanging LLRs between the
multiuser SISO and the bank of single-user APP decoders, and the
internal NMMSE-detector iteration, improving the NMMSE estimate. A
further design parameter is the choice of the initial solution
$\m^0$. Typical choices are $\m^0=\mathbf{0}$, $\m^0=\tran{\S} \r$
or $m^0_k = \mathsf{tanh} \left[\lambda_k^p/2 \right]$,
$k=1,2,...,K$, using the most recent prior LLR for user $k$.

The performance of the proposed multiuser SISO detectors within an
iterative multiuser decoder is evaluated based on numerical
examples in Section \ref{sec:NR}.

\section{Large System Performance Analysis}\label{sec:LSP}
In this section, large system analysis is considered for the
uncoded case. The BER performance of the PSPDA detector in
\eref{eqn:pspda} with uniform binary priors (i.e.,
$\lambda^p_{k}/2 = 0$) and $\m^0 =\0$ is
 investigated using an approach similar to
\cite{Kab03JPA,Tan03IT}.

Let $h_k^t = A^t\left(y_k-\sum_{j\neq k}R_{kj}m_j^t\right)$, where
$A^t =[\sigma^2+\alpha\left(1-Q^t\right)]^{-1}$. We can then
express \eref{eqn:pspda} as
\begin{equation}
    m_k^{t+1} = \omega m_k^{t} + (1-\omega)\mathsf{tanh}
    \left[h_k^t \right] =
    \sum_{\kappa = 0}^t \rho^{t-\kappa} \mathsf{tanh}\left[h_k^{\kappa}
    \right], \label{eqn:rec_m}
\end{equation}
where the recursion in \eref{eqn:pspda} has been repeatedly
applied such that,
\begin{equation}\label{eqn:rec_w}
\rho^{t-\kappa} = \left\{\begin{array}{cl}
\omega^{t-1} & \mbox{if $\kappa = 0$} \\
(1-\omega)\omega^{t-\kappa}  & \mbox{if $\kappa \neq 0$}
\end{array} \right. .
\end{equation}
The corresponding decision at iteration $t+1$ is given as
\begin{equation*}
    \hat{d}_k^{t+1} = \mathsf{sign}(m_k^{t+1})=
    \mathsf{sign}\left[
    \sum_{\kappa=0}^t \rho^{t-\kappa}\mathsf{tanh}(h_k^\kappa)\right],
\end{equation*}
and the BER at iteration $t+1$ can subsequently be determined as
\begin{eqnarray} \label{eqn:large_BER}
    P_b^{t+1} &=& \frac{1}{2}\mathsf{E}\left\{1-d_k \hat{d}_k^{t+1}
    \right\}= \frac{1}{2}\mathsf{E}\left\{1-d_k \mathsf{sign}\left[
    \sum_{\kappa=0}^t \rho^{t-\kappa}\mathsf{tanh}(h_k^\kappa)\right]
    \right\} \nn \\ &=& \frac{1}{2}\mathsf{E}\left\{1-\mathsf{sign}\left[
    \sum_{\kappa=0}^t \rho^{t-\kappa}\mathsf{tanh}(d_k h_k^\kappa)\right]
    \right\},
\end{eqnarray}

Assuming that $d_kh_k^t$ is a random variable, independently
sampled from a Gaussian distribution\footnote{This assumption
becomes increasingly valid as $K,N \rightarrow \infty$ with
$K/N=\alpha$.} with mean value $E^t$ and variance $F^{t,t}$,
respectively, and corresponding pdf $\mathsf{p}_{dh}(\beta^t)$, it
follows that the BER in \eref{eqn:large_BER} can be determined
through a $t$-fold integration as
\begin{equation*}
    P_b^{t+1} = \frac{1}{2} \int_{-\infty}^{\infty} \cdots
    \int_{-\infty}^{\infty}\left(1-\mathsf{sign}\left[
    \sum_{\kappa=0}^t \rho^{t-\kappa}\mathsf{tanh}(d_k h_k^\kappa)\right]
    \right) \prod_{\iota=0}^{t} \mathsf{p}_{dh}(\beta^{\iota})
    d\beta^{\iota}.
\end{equation*}
When $\omega = 0$, \eref{eqn:large_BER} simplifies to
\begin{eqnarray} \label{eqn:simple_BER}
    P_b^{t+1} &=& \frac{1}{2}\mathsf{E}\left\{1-\mathsf{sign}\left[
    \tanh(d_k h_k^t)\right] \right\} = \frac{1}{2}\mathsf{E}
    \left\{1-\mathsf{sign}(d_kh_k^t)\right\} \nn \\
    &=& \int_{-\infty}^{0} \mathsf{p}_{dh}(\beta^{t})
    d\beta^t = \int_{-\infty}^{-E^t/\sqrt{F^{t,t}}}\, Dz,
\end{eqnarray}
where the third equality in \eref{eqn:simple_BER} follows from
\begin{equation*}
1 - \mathsf{sign}(x) = \left\{\begin{array}{cl}
      0 & x \geq 0\\
      2 & x < 0 \\
    \end{array} \right.
\end{equation*}
and $Dz = dz\exp(-z^2/2)/\sqrt{2\pi}$. Under the assumption that
the tentative decision statistics $\{m_k^t\}$ in \eref{eqn:pspda}
converges to a fixed-point as $t \rightarrow \infty$,
$m_k^{t+1}=m_k^t=m_k$, and thus, $m_k = \mathsf{tanh}\left[h_k
\right]$. Consequently, the BER in steady-state can be determined
by \eref{eqn:simple_BER} for any weighting factor $0 \leq \omega
\leq 1$ using the steady-state distribution
$\mathsf{p}_{dh}(\beta)$ with mean value $E$ and variance $F$.

The task is therefore to derive useful recursive expressions for
$E^t$ and $F^{t,t}$. For this purpose, we define the following
parameters, $M^t$ and $Q^t$. These parameters turn out to be
closely related to $E^t$ and $F^{t,t}$.
\begin{eqnarray}\label{eqn:M1}
    M^{t+1} &=& \mathsf{E}\{d_km_k^{t+1}\} =
    \omega \mathsf{E}\{d_km_k^t\}+
    (1-\omega) \mathsf{E}\{\mathsf{tanh}(d_kh_k^{t})\}
    = \omega M^t + (1-\omega)I^t,
\end{eqnarray}
and
\begin{eqnarray} \label{eqn:Q1}
  Q^{t+1} &=& \mathsf{E}\{(m_k^{t+1})^2\} \nn \\
   &=& \omega^2 \mathsf{E}\{(m_k^t)^2\} +
(1-\omega)^2\mathsf{E}\{\mathsf{tanh}^2(d_kh_k^t)\}
+ 2\omega(1-\omega)\mathsf{E}\{m_k^t\mathsf{tanh}(h_k^t)\} \nn \\
   &=& -\omega^2Q^{t}+ 2\omega Q^{t+1,t}+ (1-\omega)^2J^t,
\end{eqnarray}
where
\begin{eqnarray*}
  I^{t} &=& \mathsf{E}\{\mathsf{tanh}(d_kh_k^{t})\}
  = \int_{-\infty}^{\infty} \mathsf{tanh}(\beta^t)
  \mathsf{p}_{dh}(\beta^{t})
    d\beta^t \nn = \int_{-\infty}^{\infty}
    \mathsf{tanh}(z\sqrt{F^{t,t}}+E^t)\,Dz, \nn \\
  J^{t} &=& \mathsf{E}\{\mathsf{tanh}^2(d_kh_k^{t})\}
  = \int_{-\infty}^{\infty} \mathsf{tanh}^2(\beta^t)
  \mathsf{p}_{dh}(\beta^{t})
    d\beta^t = \int_{-\infty}^{\infty}
    \mathsf{tanh}^2(z\sqrt{F^{t,t}}+E^t)\,Dz.
\end{eqnarray*}
The correlation $Q^{t+1,\tau}$ is given by \beqna Q^{t+1,\tau} &=&
\omega\mathsf{E}\{m_k^{t}m_k^{\tau}] +
(1-\omega)\mathsf{E}\{m^{\tau}\mathsf{tanh}(h_k^{t})\}\nn\\ &=&
\omega Q^{t,\tau} + (1-\omega) \sum_{\kappa =
0}^{\tau-1}\rho^{\tau-1-\kappa}\,\mathsf{E}\left\{\mathsf{tanh}(h_k^{t})
\mathsf{tanh}(h_k^\kappa)\right\}. \label{eqn:Q_tn}\eeqna In order
to get an expression for $Q^{t+1,\tau}$, we need to derive an
expression for
$\mathsf{E}\{\mathsf{tanh}(h_k^{t})\mathsf{tanh}(h_k^{\kappa})\}$.
We first note that $(d_kh_k^{t},d_kh_k^{\kappa})$ has a joint
Gaussian probability distribution function with
\begin{equation*}
    \mathsf{E}\left\{d_kh_k^{t},d_kh_k^{\kappa}\right\} = (E^t,E^{\kappa}),
    \hspace{3mm} \mathsf{Cov}
    \left(d_kh_k^{t},d_kh_k^{\kappa}\right)= \bmat F^{t,t} & F^{t,\kappa} \\
F^{t,\kappa} & F^{\tau,\tau} \bemat.
\end{equation*}
Rewriting $d_kh_k^{t}$ and $d_kh_k^{\kappa}$ in terms of three
independent, zero-mean, unit-variance Gaussian random variables
$\{a,b,c\}$, and the statistics above, we get \beqna d_kh_k^t =
\sqrt{F^{t,t}} \left(a\Gamma_1^{t,\kappa} +
c\Gamma_2^{t,\kappa}\right) + E^t \hspace{4mm} \text{and}
\hspace{4mm} d_kh_k^\kappa = \sqrt{F^{\kappa,\kappa}}
\left(b\Gamma_1^{t,\kappa} + c\Gamma_2^{t,\kappa}\right) +
E^\kappa, \nn\eeqna where \beqna \Gamma_1^{t,\kappa} =
\sqrt{1-\frac{F^{t,\kappa}}{\sqrt{F^{t,t}F^{\kappa, \kappa}}}}
\hspace{4mm} \text{and} \hspace{4mm} \Gamma_2^{t,\kappa} =
\sqrt{\frac{F^{t,\kappa}}{\sqrt{F^{t,t}F^{\kappa,
\kappa}}}}.\nn\eeqna It follows that \beqna
\mathsf{E}\{\mathsf{tanh}(h_k^{t})\mathsf{tanh}(h_k^{\kappa})\}
&=&
\int_{\infty}^{\infty}\int_{\infty}^{\infty}\int_{\infty}^{\infty}
\mathsf{tanh}\left[\sqrt{F^{t,t}} \left(a\Gamma_1^{t,\kappa} +
c\Gamma_2^{t,\kappa}\right) + E^t\right]\nn\\ &&
\hspace{1.5cm}\times \mathsf{tanh}\left[\sqrt{F^{\kappa,\kappa}}
\left(b\Gamma_1^{t,\kappa} + c\Gamma_2^{t,\kappa}\right) +
E^\kappa\right]\,Da\,Db\,Dc.\nn\eeqna Thus, in order to determine
$Q^{t+1}$, we need to determine the covariance  between $d_kh_k^t$
and $d_kh_k^\tau$ denoted by $F^{t,\tau}$.

In the large-system limit, the sample mean converges to the
ensemble expectation. Exploring that at stage $t$, $d_kh_k^t$ is
independently sampled, we can then determine the mean, variance
and covariance as
\begin{align*}
    E^t &= \mathsf{E}\left\{d_k h_k^t\right\} = \frac{1}{K} \sum_{k=1}^K d_k
    h_k^t, \text{ for } K \rightarrow \infty \\
    F^{t,t} &= \mathsf{Var}\left\{d_k h_k^t\right\} = \frac{1}{K} \sum_{k=1}^K
    \left(h_k^t\right)^2 - \left(E^t\right)^2, \text{ for } K \rightarrow \infty \\
    F^{t,\tau} &= \mathsf{Cov}
    \left(d_k h_k^{t},d_k h_k^{\tau}\right) = \frac{1}{K} \sum_{k=1}^K
    h_k^t h_k^{\tau} - \frac{1}{K^2} \sum_{j=1}^K \sum_{l=1}^K
    h_j^t h_l^{\tau}, \text{ for } K \rightarrow \infty
\end{align*}

Considering the correlation between $d_j$, $R_{kj}$ and $m_k^t$,
we can use methods from statistical neuro-dynamics
\cite{Ama88NN,Oka95NW} to determine $E^t$, $F^{t,t}$ and
$F^{t,\tau}$. Recently, this method has been applied to analyze
the performance of the parallel cancellation detector in
\cite{Tan03IT}. The output $h_k^t$ can be expressed as \beqna
h_k^t &=& A^t \tran{\s}_k \left(\r - \S_k \m_k^t \right) = A^t
\sum_{\mu=1}^N s_{\mu k} \left(r_{\mu} - \sum_{j \neq k} s_{\mu j}
m_j^t \right) = A^t d_k \frac{1}{\sqrt{N}}\sum_\mu z_{\mu k}^t,
\label{eqn:h_k^t}\eeqna where \beqna z_{\mu k}^t &=&
\sqrt{N}d_ks_{\mu k}\left(r_{\mu} - \sum_{j \neq k}s_{\mu j}m_j^t
\right) = \sqrt{N}d_ks_{\mu
k}r_\mu - \sqrt{N}d_ks_{\mu k}\sum_{j\neq k}s_{\mu j}m_j^t \nn\\
&=& \sqrt{N}d_ks_{\mu k}r_\mu - \omega\sqrt{N}d_ks_{\mu
k}\sum_{j\neq k}s_{\mu j}m_j^{t-1} - (1-\omega)\sqrt{N}d_ks_{\mu
k}\sum_{j\neq k}s_{\mu j}\mathsf{tanh} \left(h_j^{t-1}\right)\nn\\
&=& \omega z_{\mu k}^{t-1} + (1-\omega)\left[\sqrt{N}d_ks_{\mu
k}r_\mu - \sqrt{N}d_ks_{\mu k}\sum_{j\neq k}s_{\mu
j}\mathsf{tanh}\left(h_j^{t-1}\right)
\right].\label{eqn:z_uk}\eeqna As we aim for using
\eref{eqn:h_k^t} in determining $E^t$ and $F^{t,t}$, the
derivations are complicated by $s_{\mu j}$ and $\mathsf{tanh}
\left(h_{j}^{t-1}\right)$ being statistically dependent. To obtain
a recursive relation, the terms $\mathsf{tanh}
\left(h_j^{t-1}\right)$ are therefore expanded to separate the
dependence of $\mathsf{tanh} \left(h_{j}^{t-1}\right)$ and $s_{\mu
j}$. This can be achieved via a Taylor expansion,
$\mathsf{f}(x)\thickapprox \mathsf{f}(x_0) +
\mathsf{f}'(x_0)(x-x_0)$, as follows \beqna
\mathsf{tanh}\left(h_{j}^{t-1}\right) &\thickapprox&
\mathsf{tanh}\left(h_{\mu j}^{t-1}\right) +
\mathsf{sech}^2\left(h_{\mu
j}^{t-1}\right) \left(h_{j}^{t-1}-h_{\mu j}^{t-1} \right) \nn \\
&=& \mathsf{tanh}\left(h_{\mu j}^{t-1}\right) +
\mathsf{sech}^2\left(h_{\mu j}^{t-1}\right)
A^{t-1}\left[\sum_{i\neq j}s_{\mu i} s_{\mu
j}\left(d_i-m_i^{t-1}\right) + s_{\mu j}n_\mu\right],\nn\eeqna
where $h_{\mu j}^{t-1}$ is chosen such that it contains no terms
with $s_{\mu j}$, \beqna h_{\mu j}^{t-1} &=& A^{t-1}\left[d_k +
\sum_{\nu\neq\mu}\sum_{j\neq k}s_{\nu k} s_{\nu
j}\left(d_j-m_j^{t-1}\right) + \sum_{\nu\neq\mu}s_{\nu
k}n_\nu\right], \nn\eeqna and $\mathsf{sech}(x) =
\mathsf{cosh}^{-1}(x)$. The term $\sum_{j\neq k}s_{\mu
j}\mathsf{tanh}\left(h_j^{t-1}\right)$ in \eref{eqn:z_uk} can now
be expressed as \beqna \sum_{j\neq k}s_{\mu
j}\mathsf{tanh}\left(h_j^{t-1}\right) &\thickapprox& \sum_{j\neq
k}s_{\mu j} \mathsf{tanh}\left(h_{\mu j}^{t-1}\right) \nn\\&&
\hspace{-1cm} + \sum_{j\neq k}s_{\mu j}
\mathsf{sech}^2\left(h_{\mu j}^{t-1}\right)
A^{t-1}\left[\sum_{i\neq j}s_{\mu i} s_{\mu
j}\left(d_i-m_i^{t-1}\right) + s_{\mu j}n_\mu\right]\label{eqn:step}\\
&\thickapprox&   \sum_{j\neq k}s_{\mu j} \mathsf{tanh}\left(h_{\mu
k}^{t-1}\right) + \alpha U^t A^{t-1}\left[ r_\mu - \sum_{i}s_{\mu
i} m_i^{t-1}\right], \label{eqn:sfh}\eeqna where \beqna U^t &=&
\frac{1}{K} \sum_j \mathsf{sech}^2\left(h_{j}^{t-1}\right).
\label{eqn:U}\eeqna In the second term in the step from
\eref{eqn:step} to \eref{eqn:sfh}, the two summations have been
extended over all $j$ and $i$, respectively, simplifying the
derivations below. In the large system limit, these few extra
terms included in the summations do not affect the final results.

Substitute \eref{eqn:sfh} into \eref{eqn:z_uk}, we have \beqna
z_{\mu k}^t &\thickapprox& \omega z_{\mu k}^{t-1} +
(1-\omega)\left[\sqrt{N}d_ks_{\mu k}r_\mu - \sqrt{N}d_ks_{\mu
k}\sum_{j\neq k}s_{\mu j}\mathsf{tanh}\left(h_{\mu j}^{t-1}\right)
\right.\nn\\ &&\left. \hspace{4cm} -\alpha U^t
A^{t-1}\left(\sqrt{N}d_ks_{\mu k} r_\mu - \sqrt{N}d_ks_{\mu
k}\sum_{i\neq k}s_{\mu i} m_i^{t-1}\right)\right] \nn\\
&=& \omega z_{\mu k}^{t-1} + (1-\omega)\left[\bar{z}_{\mu k}^t -
\alpha U^t A^{t-1} z_{\mu k}^{t-1} \right],\label{eqn:z_uk1}\eeqna
where \beqna \bar{z}_{\mu k}^t &=& \sqrt{N}d_ks_{\mu k}r_\mu -
\sqrt{N}d_ks_{\mu k}\sum_{j\neq k}s_{\mu
j}\mathsf{tanh}\left(h_{\mu j}^{t-1}\right). \nn\eeqna With $\m^0
= \0$, we can find $z_{\mu k}^0 = \sqrt{N}d_ks_{\mu k} r_\mu$, and
then using \eref{eqn:z_uk1} recursively, we can determine $z_{\mu
k}^t$. Letting $B^t = \mathsf{E}\left\{\sqrt{N}z_{\mu
k}^t\right\}$, we can also use \eref{eqn:z_uk1} to arrive at the
following recursive relationship \beqna B^t &=& \omega B^{t-1} +
(1-\omega)\left[1-\alpha U^{t}A^{t-1}B^{t-1}\right],
\label{eqn:B}\eeqna where $\mathsf{E}\left\{\sqrt{N}\bar{z}_{\mu
k}^t\right\} = 1$, since $s_{\mu j}$ and
$\mathsf{tanh}\left(h_{\mu j}^{t-1}\right)$ are approximately
independent.

Finally, using \eref{eqn:z_uk1}, the covariance of $z_{\mu k}^t$
and $z_{\mu k}^\tau$ is given as \beqna C^{t,\tau} &=&
\mathsf{Cov}\left\{z_{\mu k}^t z_{\mu k}^\tau\right\} \nn\\ &=&
\omega^2 C^{t-1,\tau-1} +
\omega(1-\omega)\left[\mathsf{E}\left\{z_{\mu k}^{t-1}\bar{z}_{\mu
k}^\tau\right\} - \alpha U^{\tau}A^{\tau-1}C^{t-1,\tau-1} \right]
\nn\\ && + \omega(1-\omega)\left[\mathsf{E}\left\{z_{\mu
k}^{\tau-1}\bar{z}_{\mu k}^t\right\} - \alpha
U^{t}A^{t-1}C^{t-1,\tau-1} \right] \nn\\ && + (1-\omega)^2
\left[V^{t,\tau} - \alpha U^\tau A^{\tau-1}\mathsf{E}\left\{z_{\mu
k}^{\tau-1}\bar{z}_{\mu k}^t\right\} \right. \nn\\ &&
\hspace{2.5cm}\left. - \alpha U^t A^{t-1}\mathsf{E}\left\{z_{\mu
k}^{t-1}\bar{z}_{\mu k}^\tau\right\} + \alpha^2 U^t A^{t-1} U^\tau
A^{\tau-1}C^{t-1,\tau-1} \right],\label{eqn:C}\eeqna where \beqna
V^{t,\tau} &=& \mathsf{E}\left\{\bar{z}_{\mu k}^{t}\bar{z}_{\mu
k}^\tau\right\} = \alpha + \sigma^2 - \alpha I^{t-1} - \alpha
I^{\tau-1} + \alpha \mathsf{E}\left\{\mathsf{f}\left(h_{\mu
j}^{t-1}\right)\mathsf{f}\left(h_{\mu j}^{\tau-1}\right)
\right\}.\label{eqn:V} \eeqna The two remaining terms
$\mathsf{E}\left\{z_{\mu k}^{t-1}\bar{z}_{\mu k}^\tau\right\}$ and
$\mathsf{E}\left\{z_{\mu k}^{\tau-1}\bar{z}_{\mu k}^t\right\}$ can
be determined recursively from $\mathsf{E}\left\{z_{\mu
k}^{0}\bar{z}_{\mu k}^t\right\}$. These derivations are
straightforward and have been omitted to save space.

Now we have all the terms required to determine the mean $E^t$ and
the covariance $F^{t,\tau}$. Since $h_k^t = A^t d_k \sum_\mu
z_{\mu k}^t/\sqrt{N}$, it follows that $E^t$ and $F^{t,\tau}$ are
given by \beqna E^t &=& \mathsf{E}\left\{d_k h_k^t \right\}  =
\mathsf{E}\left\{ A^t\sum_\mu z_{\mu k}^t/\sqrt{N}\right\} =
A^tB^t, \label{eqn:E2}\eeqna and \beqna F^{t,\tau} &=&
\mathsf{Cov}\left\{d_k h_k^t d_k h_k^{\tau}\right\} =
\mathsf{Cov}\left\{h_k^th_k^\tau\right\} = A^t A^{\tau}
\mathsf{Cov}\left\{\sum_\mu \sum_\nu z_{\mu k}^tz_{\mu k}^\tau
\right\}/N  \nn\\
&=& A^tA^\tau \mathsf{Cov}\left\{z_{\mu k}^tz_{\mu k}^\tau\right\}
= A^tA^\tau C^{t,\tau} ,\label{eqn:F2}\eeqna respectively. Note
that for $\omega = 0$ and as $U^t \rightarrow 0$, \eref{eqn:E2}
and \eref{eqn:F2} tend to \beqna E^{t} &=&
\frac{1}{\sigma^2+\alpha(1-Q^t)},\label{eqn:E1}\\
F^{t,t} &=&
\frac{\alpha(1-2M^t+Q^t)+\sigma^2}{[\sigma^2+\alpha(1-Q^t)]^2},
\label{eqn:F1}
\eeqna respectively. It has been observed that $U^t \rightarrow 0$
when $E^t$ and $F^{t,t}$ increase. More importantly, equations
\eref{eqn:M1}, \eref{eqn:Q1}, \eref{eqn:E1} and \eref{eqn:F1} are
identical to the fixed point iterations of the saddle point
equations found by the replica method analysis for optimal
detection \cite{Tan02IT}. Hence, the expressions obtained above
link the simplified PDA detector to the replica analysis of the
equilibrium state presented in \cite{Tan02IT} for uniform binary
priors. Based on the large system analysis in this section, we
conclude that the simplified PDA detector approaches the
performance of the optimal detector as $K$ and $N$ grows large
with $\alpha=K/N$ and transmission is conducted at a sufficiently
large $E_b/N_0$.

Finally, under the assumption that the tentative decisions
$\{m_k^t\}$ in \eref{eqn:pspda} converge as $t \rightarrow
\infty$, we can regard all quantities as being independent of
subscripts $t$ and $\tau$. Following from \eref{eqn:M1},
\eref{eqn:Q1}, \eref{eqn:U}, \eref{eqn:B}-\eref{eqn:F2}, the
equilibrium conditions are then given by
\beqna M &=& I = \int \mathsf{tanh}\left(z\sqrt{F} + E\right)\,Dz \label{eqn:M_e}\\
Q &=& J = \int \mathsf{tanh}^2\left(z\sqrt{F} + E\right)\,Dz \label{eqn:Q_e}\\
U &=& \int \mathsf{sech}^2\left(z\sqrt{F} + E\right)\,Dz
\label{eqn:U_e}\\ A &=& \frac{1}{\sigma^2 +
\alpha(1-Q)}\label{eqn:A_e}\\ E &=& \frac{A}{1+\alpha UA}
\label{eqn:E_e}\\ F &=&
\frac{A^2[\sigma^2+\alpha(1-2M+Q)]}{(1+\alpha UA)^2}
\label{eqn:F_e}\eeqna With initial values for $M$, $Q$ and $U$, we
can then recursively find the steady-state solution to the above
equations, leading to a numerical approach determining the
large-system $E$ and $F$, and thus the corresponding large-system
BER performance.

\section{Numerical Results}\label{sec:NR}
In this section we illustrate the results above through numerical
examples.
\begin{figure}[hbtp]
\vspace{.01cm}

  \begin{center}
    \setlength{\unitlength}{.625mm}\small
    \subfigure[$\alpha = 0.25$.\label{fig:f1a}]{
      \begin{picture}(120,103)(5,0)
        \put(0,0){\includegraphics*[width=80mm]{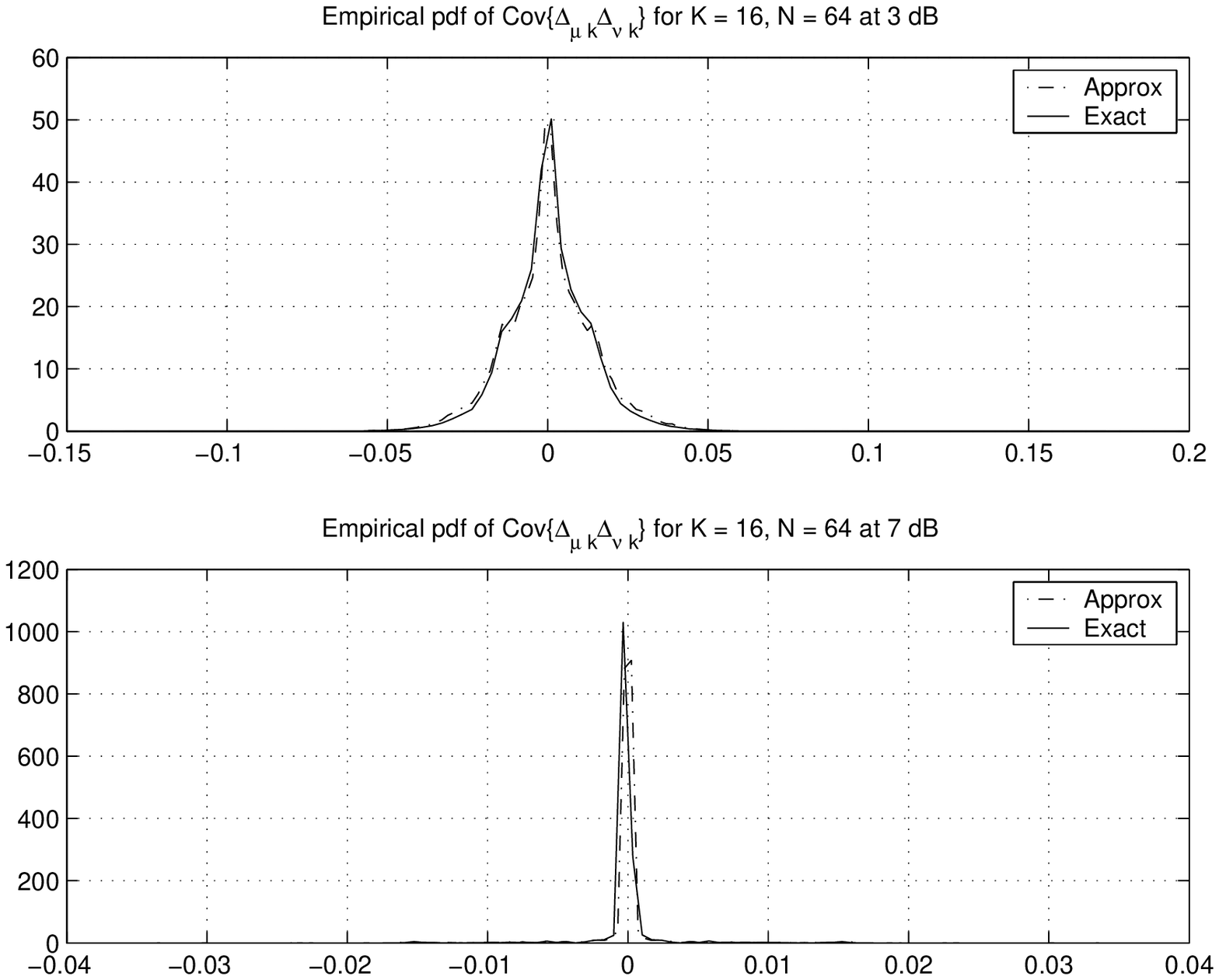}}
      \end{picture}
    }
    \hspace{5mm}
    \subfigure[$\alpha = 1$. \label{fig:f1b}]{
      \begin{picture}(120,103)(5,0)
        \put(0,0){\includegraphics*[width=80mm]{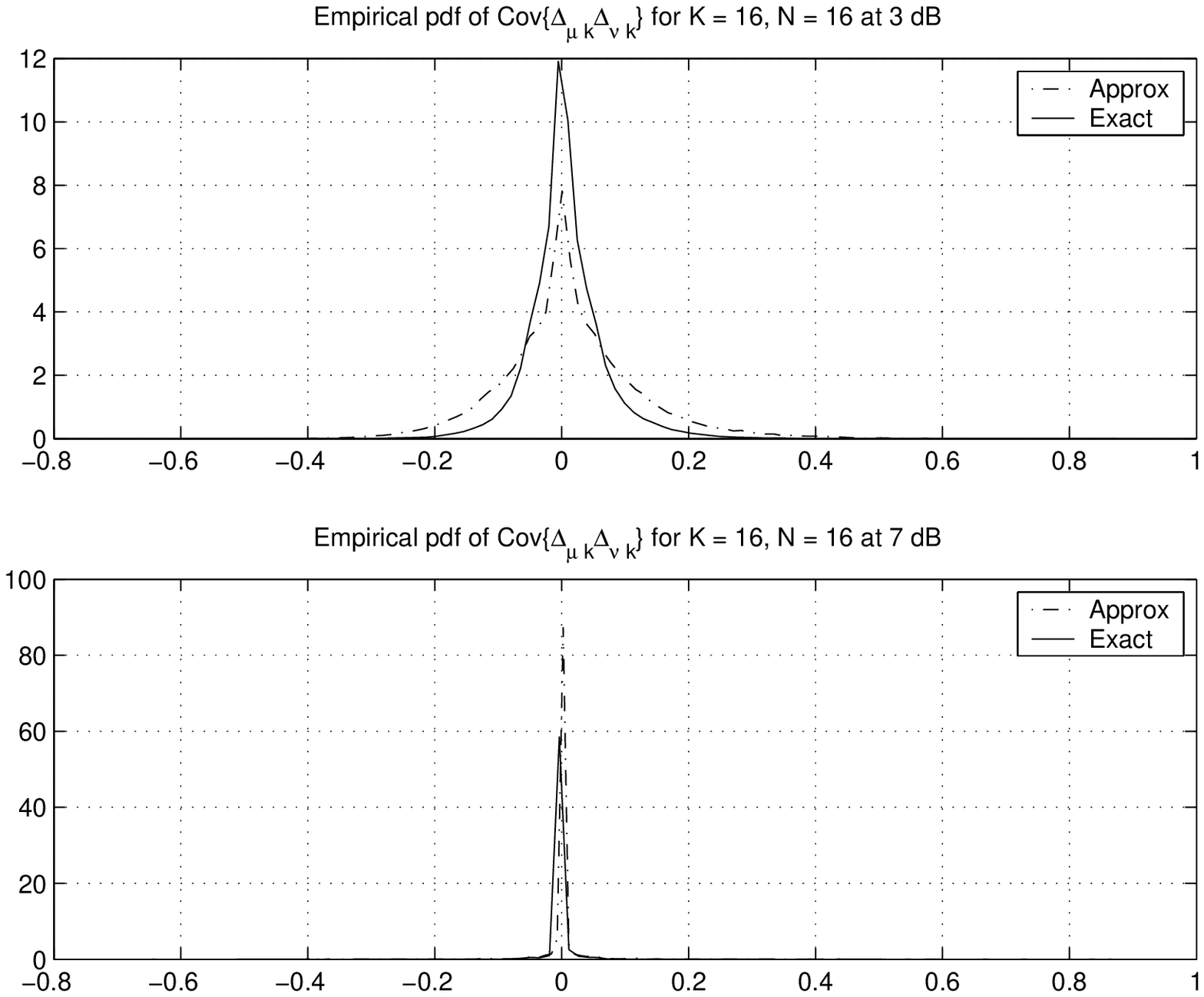}}
      \end{picture}
    }
  \end{center}
  \caption{Empirical pdf of $\mathsf{Cov}\{\Delta_{\mu k}\Delta_{\nu k}\}$
  for different $\alpha$.}
  \label{fig:f1}
\end{figure}
First, the empirical pdfs of $\mathsf{Cov}\{\Delta_{\mu
k}\Delta_{\nu k}\}$ in \eref{eqn:cov1} is investigated. Figure
\ref{fig:f1} shows the empirical pdf with and without the second
term in \eref{eqn:cov1}. For a lightly loaded system ($\alpha =
0.25$), omitting the second term has only a minor effect on the pdf
as seen in Figure \ref{fig:f1a}. The difference is more pronounced
when the load increases to 1, as shown in Figure \ref{fig:f1b}.
Here, we can only simulate systems with a small number of users
($K=16$) due to the computational complexity of determining the
optimal marginal posterior-mode mean values $m_k$. We expect the
difference between the exact and the approximation to be reduced
when $K$ and $N$ increase.

\begin{figure}[hbtp]
  \begin{center}
    \setlength{\unitlength}{.625mm}\small
    \subfigure[Comparison of replica analysis (RA), statistical neurodynamics
    (SN) and simulation results for $E_b/N_0 = 6,7,8,9$ dB, $K = 512$, and $\alpha = 0.1$.
    \label{fig:f3a}]{
      \begin{picture}(120,103)(5,0)
        \put(0,0){\includegraphics*[width=80mm]{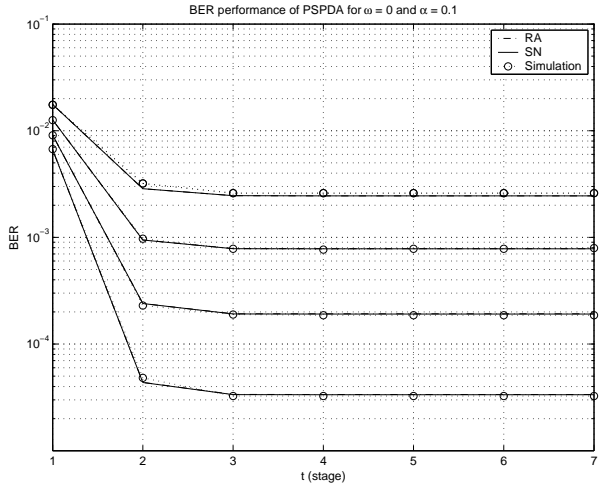}}
      \end{picture}
    }
    \hspace{5mm}
    \subfigure[Comparison of replica analysis (RA), statistical neurodynamics
    (SN) and simulation results for $E_b/N_0 = 6,7,8,9$ dB, $K = 512$, and $\alpha = 0.5$.
    \label{fig:f3b}]{
      \begin{picture}(120,103)(5,0)
        \put(0,0){\includegraphics*[width=80mm]{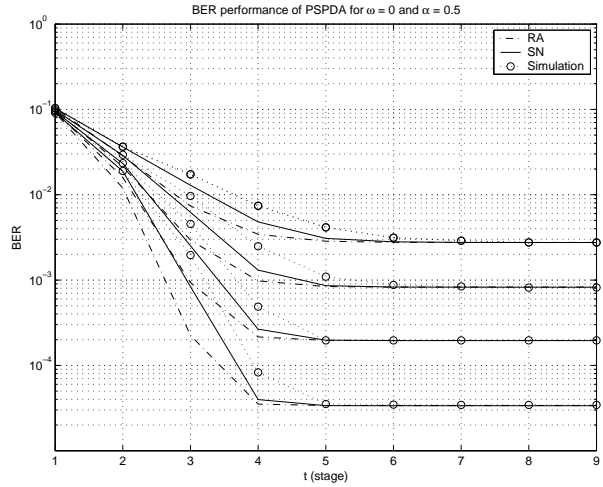}}
      \end{picture}
    }
  \end{center}
  \caption{BER approximation.}
  \label{fig:f3}
\end{figure}

Now we consider the large system BER estimates derived for the
PSPDA ($\omega = 0$) through the replica analysis (RA) and
statistical neurodynamics (SN) approach in Figure \ref{fig:f3}.
The BER estimates for the SN approach are obtained from iterating
\eref{eqn:M1}, \eref{eqn:Q1}, \eref{eqn:E2} and \eref{eqn:F2},
whilst the BER estimates for the RA approach are obtained from
iterating \eref{eqn:M1}, \eref{eqn:Q1}, \eref{eqn:E1} and
\eref{eqn:F1}. When the load is small ($\alpha = 0.1$ in Figure
\ref{fig:f3a}), the simulated BER performance coincide with those
estimates from the SN and RA approach. As the load increases to
$0.5$ in Figure \ref{fig:f3b}, the simulated BER performance do
not follow the SN and RA approach in the first few stages. But it
does converge to the estimates given by the SN and RA approach.

\begin{figure}[hbtp]
  \begin{center}
    \setlength{\unitlength}{.625mm}\small
    \subfigure[Small system.\label{fig:f4a}]{
      \begin{picture}(120,103)(5,0)
        \put(0,0){\includegraphics*[width=80mm]{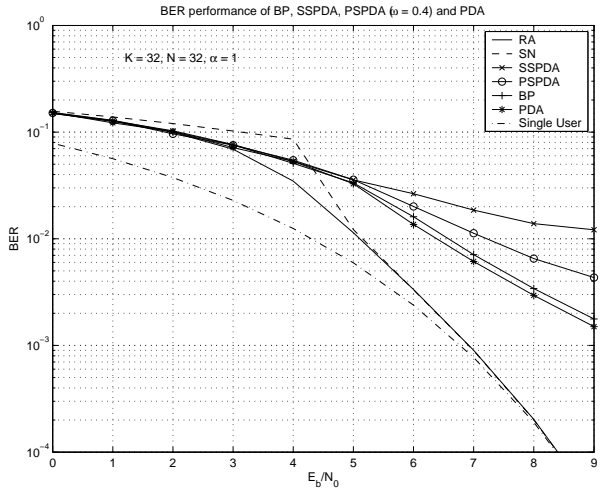}}
      \end{picture}
    }
    \hspace{5mm}
    \subfigure[Large system. \label{fig:f4b}]{
      \begin{picture}(120,103)(5,0)
        \put(0,0){\includegraphics*[width=80mm]{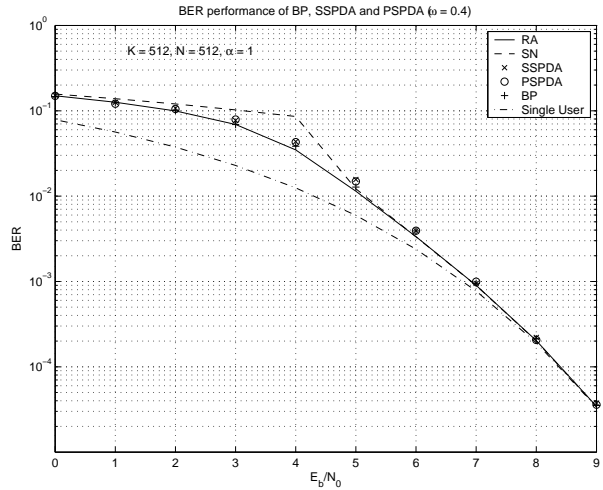}}
      \end{picture}
    }
  \end{center}
  \caption{Comparison of BER performance of the BP, PSPDA,  SSPDA
            and PDA detectors for uncoded systems with uniform prior probabilities.}
  \label{fig:f4}
\end{figure}

In Figure \ref{fig:f4}, the BER performance of BP \cite{Kab03JPA},
PSPDA ($\omega = 0.4$) and the SSPDA detectors is compared to the
RA and SN predicted performance for an uncoded CDMA system with
$\alpha=1$.
Convergence is considered achieved when
$\max|m_k^t-m_k^{t-1}|<10^{-3}$ or the number of iterations has
exceeded $100$. Table \ref{tab:t1} shows the average number of
stages required for convergence. As $E_b/N_0$ increases, the SSPDA
detector converges faster and hence requires the least computational
complexity.
\begin{table}
\begin{center}
\begin{tabular}{c|cccccccccc} \hline
$E_b/N_0/dB$  &  0 &  1 &  2 &  3 & 4 &  5 & 6 &  7 & 8 & 9 \\
\hline
SSPDA & 12.5 & 23.6 & 77.9 & 62.4 & 48.9 & 27.1 & 14.4 & 8.9 & 6.8 & 5.8 \\
PSPDA & 31.3 & 58.0 & 99.0 & 99.0 & 88.1 & 62.4 & 36.6 & 24.1 & 19.1
& 17.1
\\ BP & 16.4 & 20.3 & 27.3 & 39.1 & 50.0 & 39.6 & 23.1 & 14.4 & 10.0 & 7.7 \\ \hline
\end{tabular}
\end{center}
\caption{Average number of stages required for convergence for
$K=512$ and $\alpha=1$.} \label{tab:t1}
\end{table}
As the load increases to 1, simple iterations of \eref{eqn:M1},
\eref{eqn:Q1}, \eref{eqn:E2} and \eref{eqn:F2} do not yield the
desired BER estimates for the SN approach as it get attracted to
fixed points which yield poorer BER performance. The estimates from
SN approach are obtained by searching fixed points for the nonlinear
equilibrium (\eref{eqn:M_e} - \eref{eqn:F_e}) which minimize the BER
for each $E_b/N_0$. In Figure \ref{fig:f4a} it is observed, as
expected, that for a small system ($K=32$), the BP, the PSPDA and
the SSPDA detectors do not attain the BER performance predicted by
the RA. At large $E_b/N_0$, these detectors fail to provide a useful
level of performance. In contrast, when the number of users is large
($K = 512$), the BER performance of both the BP, PSPDA and SSPDA
detectors coincide with the prediction of RA as in Figure
\ref{fig:f4b}. It is also noted that the serial SSPDA converges
faster than the BP detector, which is implemented in parallel, while
the PSPDA detector converges slower than the BP detector.

\begin{figure}[hbtp]
  \begin{center}
    \setlength{\unitlength}{.625mm}\small
    \subfigure[PIC and PDA.\label{fig:f5a}]{
      \begin{picture}(120,103)(5,0)
        \put(0,0){\includegraphics*[width=80mm]{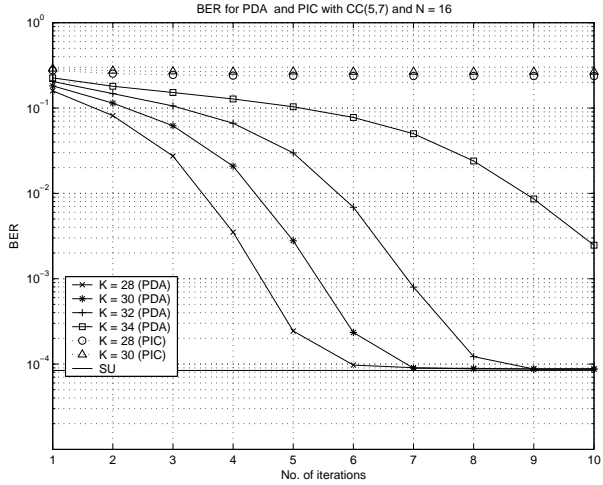}}
      \end{picture}
    }
    \hspace{5mm}
    \subfigure[BP, MIC and SSPDA. \label{fig:f5b}]{
      \begin{picture}(120,103)(5,0)
        \put(0,0){\includegraphics*[width=80mm]{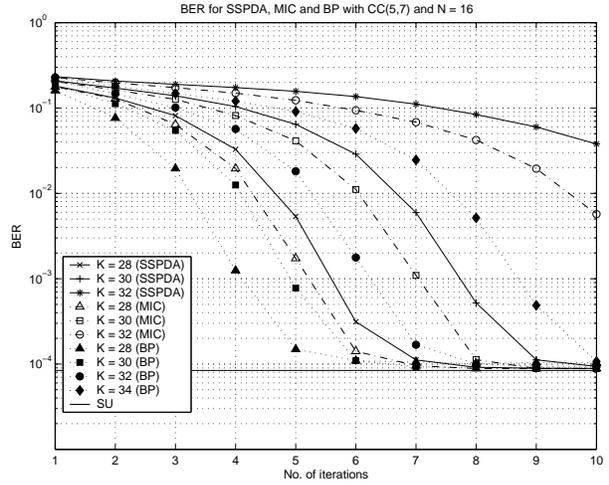}}
      \end{picture}
    }
  \end{center}
  \caption{Comparison of BER performance of the PDA, PIC, SSPDA, MIC and BP
             detectors for coded systems.}
  \label{fig:f5}
\end{figure}

In Figure \ref{fig:f5}, we compare the BER performance of the PDA
\cite{Tan03ISIT}, the parallel interference canceller (PIC) in
\cite{Tar97ISIT,Sti03VT}, the serial SSPDA \eref{eqn:sspda}, the
serial MIC \eref{eqn:mkic} and the BP detector \cite{Kab03JPA} in
a coded CDMA system where each user applied a $(5,7)$
convolutional code, the processing gain is $N = 16$, the
interleaver size is $1000$ information bits per user and iterative
multiuser detection is done as in
\cite{Wan99TC,Cai03IT,Tan03ISIT}.
The SSPDA, BP and MIC detectors are implemented with $3$ stages
each. The BP detector converges faster than the SSPDA and MIC
detectors. Since it is a small system, the MIC detector is expected
to perform better than the SSPDA detector, which is confirmed in
Figure \ref{fig:f5}, where the MIC detector approaches single-user
performance faster than the SSPDA detector. It is noteworthy that
the two additional stages of the detectors do improve the BER
performance. For $K = 28$, both the MIC, BP and SSPDA detectors
require $7$ iterations of message passing, respectively, to approach
single-user performance. The PDA detector also achieves single-user
performance with 6 iterations, but is more computational intensive.
However, it converges slower than the BP detector when the number of
users increases beyond $30$.

\section{Conclusions}\label{sec:Con}

In this paper we have used a multivariate Gaussian approximation
of the MAI to obtain a nonlinear MMSE estimate of the transmitted
bits in a multiuser system. The assumption that the MAI is a
multivariate Gaussian random variable leads to approximating
expression of the marginal posterior-mode identical to those
describing the probabilistic data association detector. Thus, the
nonlinear MMSE framework provides an alternative justification for
the PDA detector structure. A simplified PDA detector is found
through diagonalization of a matrix inversion and recognized as
having the same structure as previously suggested soft
cancellation schemes. This simplified structure lends itself to
large system analysis which is found to be closely related to the
replica method analysis for the optimal detector, and it follows
that the simplified PDA has the same predicted large system
performance as the optimal detector. As the PDA-based detectors
can output estimates of extrinsic probabilities directly, they are
well suited for iterative multiuser decoding and found to provide
single user performance at high loads. In a coded systems, it is
noted that the additional stages of the simplified PDA do improve
the BER performance, in contrast to traditional interference
cancellation.

\vspace{1cm} \noindent {\it Acknowledgement} \\
The authors would like to thank Prof. Toshiyuki Tanaka at Tokyo
Metropolitan University for helpful discussion and providing the
preprint of \cite{Tan03IT}.

\bibliographystyle{IEEEtr}

\end{document}